\documentclass[conference]{IEEEtran}
\IEEEoverridecommandlockouts
\usepackage[utf8]{inputenc}
\usepackage{graphicx}
\usepackage{amsmath, amssymb, amsfonts}
\usepackage{booktabs}
\usepackage{tabularx}
\usepackage{float}
\usepackage{multirow}
\usepackage{subcaption}
\usepackage{enumitem}
\usepackage{cite}
\usepackage{xcolor}
\usepackage{tikz}
\usetikzlibrary{positioning}
\usepackage{hyperref}
\usepackage{comment}
\usepackage{todonotes}
\hypersetup{colorlinks=true, linkcolor=blue, citecolor=blue, urlcolor=blue}

\title{Consensus-Aware AV Behavior: Trade-offs Between Safety, Interaction, and Performance in Mixed Urban Traffic}

\author{\IEEEauthorblockN{Mohammad Elayan}
\IEEEauthorblockA{\textit{Dept. of Civil \& Environmental Engineering} \\
\textit{University of Nebraska-Lincoln}\\
melayan2@huskers.unl.edu}
\and
\IEEEauthorblockN{Wissam Kontar$^*$}
\IEEEauthorblockA{\textit{Dept. of Civil \& Environmental Engineering} \\
\textit{University of Nebraska-Lincoln}\\
wkontar2@unl.edu}
}

\begin{document}

\maketitle

\begin{abstract} 

Transportation systems have long been shaped by complexity and heterogeneity, driven by the interdependency of agent actions and traffic outcomes. The deployment of automated vehicles (AVs) in such systems introduces a new challenge: achieving consensus across safety, interaction quality, and traffic performance. In this work, we position consensus as a fundamental property of the traffic system and aim to quantify it. We use high-resolution trajectory data from the Third Generation Simulation (TGSIM) dataset to empirically analyze AV and human-driven vehicle (HDV) behavior at a signalized urban intersection and around vulnerable road users (VRUs). Key metrics, including Time-to-Collision (TTC), Post-Encroachment Time (PET), deceleration patterns, headways, and string stability, are evaluated across the three performance dimensions. Results show that full consensus across safety, interaction, and performance is rare, with only 1.63\% of AV-VRU interaction frames meeting all three conditions. These findings highlight the need for AV models that explicitly balance multi-dimensional performance in mixed-traffic environments. 

\emph{Full reproducibility is supported via our open-source codebase on \href{https://github.com/wissamkontar/Consensus-AV-Analysis}{GitHub}.}


\end{abstract}


\section{Introduction}

Complexity and heterogeneity have long been defining characteristics of transportation systems, shaped by the interdependent actions of transportation agents, their behavioral differences, and the uniqueness of the transportation environment. With the increasing adoption of automated vehicles (AVs), and driving automation capabilities, comes a new layer of complexity related to \textbf{``consensus"} in the traffic environment. Here, consensus is a multi-dimension challenge across safety of the transportation agent, interactions with other transportation agents, and overall performance of the traffic. Central to this challenge is the diverging nature of these dimensions. For instance, a highly conservative maneuver may enhance safety but disrupt traffic flow, whereas a throughput-maximizing action may compromise interaction quality or increase conflict with vulnerable road users (VRUs). 

A commonality within the current modeling architecture of AVs is treating consensus as an ancillary input, rather than a fundamental property of the traffic environment. As such, we aim in this work to shift the focus towards understanding the nature of this consensus, and conceptualizing it across three interdependent dimensions: \textbf{safety}, \textbf{interaction}, and \textbf{traffic performance}. The goal is to identify behavioral strategies where objectives across all three dimensions can coexist without major compromise. Ultimately, this leads to the design of an AV agent that makes the AV-traffic system more than the sum of its part, rather an agent of consensus. 

We take an empirical data-driven approach and analyze high-resolution trajectory data from a signalized urban intersection, specifically the TGSIM data \cite{fhwa2025tgsim}. We quantify multiple surrogate and behavioral metrics for AVs and human-driven vehicles (HDVs), including Time-to-Collision (TTC), Post-Encroachment Time (PET), deceleration patterns, headways, and string stability. These measures are not only compared across vehicle types but also evaluated in terms of how well they align or conflict across the three performance dimensions.

Our findings show that out of all interaction frames analyzed (specifically between AVs and VRUs), only 1.63\% met all three consensus conditions simultaneously, 27.65\% satisfied two conditions, and 70.71\% met at most one or none. This indicates that full consensus across safety, interaction, and performance remains a rare reality.

\begin{figure}[H]
\centering
\resizebox{\columnwidth}{!}{%
\begin{tikzpicture}[node distance=0.8cm and 1.5cm, auto, every node/.style={align=center, font=\footnotesize}]
\node[draw, rectangle, fill=blue!10] (input) {High-Resolution \\ Trajectory Data};

\node[draw, rectangle, fill=green!10, below left=of input] (safety) {Safety Metrics\\(TTC, PET, Headway)};
\node[draw, rectangle, fill=orange!10, below=of input] (interaction) {Interaction Metrics\\(Deceleration, Speed, Hesitation)};
\node[draw, rectangle, fill=purple!10, below right=of input] (performance) {Traffic Performance\\(Headway, Disruption, Clearance)};
\node[draw, rectangle, fill=gray!10, below=1.5cm of interaction] (compare) {Metric Comparison \\ Across AVs and HDVs};
\node[draw, rectangle, fill=yellow!20, below=1.5cm of compare] (reward) {Consensus \\ (Safety \& Efficiency Alignment)};

\draw[->, thick] (input) -- (safety);
\draw[->, thick] (input) -- (interaction);
\draw[->, thick] (input) -- (performance);
\draw[->, thick] (safety) -- (compare);
\draw[->, thick] (interaction) -- (compare);
\draw[->, thick] (performance) -- (compare);
\draw[->, thick] (compare) -- (reward);
\end{tikzpicture}
} 
\caption{Framework for deriving  consensus trajectory-derived safety, interaction, and performance metrics.}
\label{fig:flowchart_framework}
\end{figure}
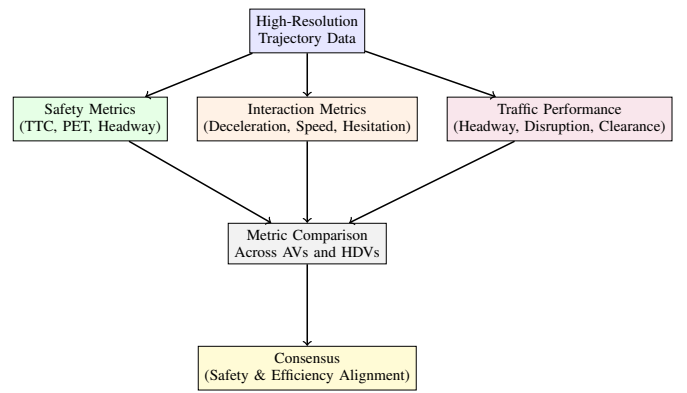

Figure~\ref{fig:flowchart_framework} illustrates our proposed framework.

\section{Literature Review}

Recent studies have demonstrated the distinct behavioral patterns of AVs compared to HDVs, particularly in interactions with VRUs. AVs tend to maintain larger safety margins and exhibit higher Time-to-Collision (TTC) and Post-Encroachment Time (PET) values, which are often interpreted as signs of more conservative, safety-conscious behavior \cite{Saeed_Rahmani_2024, Sunny_Singh_2023}. This conservatism is especially pronounced in unsignalized intersections and in response to occluded or unpredictable pedestrians, where AVs often employ proportional braking or maintain larger gaps \cite{Zou2023-dn, Hammami2024-ul}.

However, these surrogate safety metrics (SSMs) like TTC and PET, while useful, have limitations when interpreted in mixed traffic environments. Their thresholds are traditionally calibrated for human driving behavior and do not account for AVs' rapid response capabilities or rule-based operations \cite{Tafidis2023-cn, Zhang2022-jo}. For instance, shorter TTC values may be flagged as unsafe despite being within operational limits for AVs, particularly in low-speed, predictable scenarios. Furthermore, PET thresholds do not always reflect the nuances of VRU behavior or vehicle intent, leading to misclassifications in risk analysis \cite{Xhoxhi2024-zs}.

Another critical issue in the interpretation of SSMs lies in the inherent complexity of AV-VRU interactions. AVs rely heavily on environmental predictability and often assume cooperative pedestrian behavior, which may not align with real-world pedestrian decision-making, especially under ambiguity or in dense urban environments \cite{Rezwana2025-pt, Lanzaro2023-gl}. This misalignment can result in hesitance or uncoordinated movements on both sides, undermining interaction quality despite seemingly safe TTC/PET values. Moreover, while AVs maintain greater longitudinal spacing during exit, their proximity to VRUs during conflict resolution appears more frequent, potentially due to overreliance on internal models of pedestrian predictability.

The literature also highlights that AVs lack adaptive social signaling, such as eye contact or informal gestures, which HDVs frequently rely on to navigate ambiguous scenarios. This absence can reduce pedestrian trust and lead to longer gap acceptance or erratic pedestrian behaviors—dynamics not fully captured by current safety metrics \cite{Izquierdo2024-yu, Lau2024-ex}. External Human-Machine Interfaces (eHMIs) have shown some promise in addressing this gap, but widespread implementation and contextual adaptation remain limited.

Beyond surrogate safety, traffic stability and traffic performance are essential in understanding AV behavior in their influence on traffic flow dynamics. While AV-led platoons benefit from smoother disturbance attenuation and reduced stop-and-go propagation under certain control regimes \cite{Luo2024-bk, Hu2021-kq}, they also introduce greater spacing and slower accelerations in mixed platoons, particularly at signal exits, contributing to potential throughput penalties. HDV traffic, although notoriously unstable, have demonstrated more responsive behavior under pressure, attenuating follower acceleration noise more effectively in some contexts \cite{Das2024-tr}. In prior work, \cite{kontar2021multi}, we also show how AVs with different control logic (and constraints) can have different traffic wave disturbance propagation properties. 

Despite growing recognition of the multi-dimensional nature of AV performance, most evaluation frameworks remain fragmented. There is a lack of consensus-based methodologies that integrate safety, interaction quality, and traffic performance into a unified decision-making or training paradigm. Attempts such as the SAFR-AV platform or risk factor (RF) metrics offer partial solutions \cite{Pathrudkar2023-xo, Xhoxhi2024-zs}, but they are either simulation-bound or focused narrowly on specific use-cases. A consensus framework that seeks zones of agreement across various performance metrics, may offer a more balanced way forward, particularly for AV behavior design in mixed traffic settings.

While consensus-based strategies offer theoretical benefits for coordinating AVs, most remain untested in the complex dynamics of real-world traffic. Prior work has explored signal-free intersection protocols \cite{Difilippo2022-sm}, distributed trajectory planning \cite{Mirheli2019-ds}, and robust consensus under delays \cite{Loria2024-bp}, but often in controlled or simulation-based settings. Event-triggered control \cite{Liu2023-qw} and safety-preserving barrier functions \cite{Niu2023-sm} enhance feasibility, yet their practical integration with diverse road users is still limited. This paper contributes to the consensus-aware AV literature by empirically analyzing AV behavior during under mixed traffic conditions in urban settings. We ground our analysis in real trajectory data, to explore how AV decisions reflect or diverge from consensus objectives. Consequently, this work offers a data-driven look at the operational realities and contradictions within consensus-aware AV development. 

In summary, while AVs generally appear more conservative and yield higher safety margins, these metrics alone may obscure challenges in real-world interactions and systemic performance trade-offs. As AV deployment expands, especially in pedestrian-rich areas, the ability to evaluate and train behaviors across multiple interacting dimensions will be crucial to building both trust and efficiency.

\section{Dataset Description}

This study uses the \textit{TGSIM Foggy Bottom Trajectories} dataset, published in January 2025 as part of the Third Generation Simulation (TGSIM) project. The dataset captures detailed multimodal trajectories at 4 signalized urban intersections in Washington, D.C., recorded by 12 overhead 4K cameras \cite{fhwa2025tgsim}.

The \textit{TGSIM Foggy Bottom} dataset provides high-resolution trajectories for a wide range of road users in a complex urban intersection environment. Each observation includes unique agent identifiers, timestamps, smoothed position coordinates (in meters), lane assignments, and physical dimensions. Motion data are provided as \(x\) and \(y\) components of velocity and acceleration. The dataset captures a diverse mix of road users, including pedestrians, cyclists, scooter riders, buses, trucks, and both human-driven and automated passenger vehicles. 

The dataset includes 28 unique AV IDs. Of these, 14 AVs completed full trajectories covering all four intersections, while the remaining 14 exhibited partial trajectories. The AV trajectories are limited to four consecutive left-turn maneuvers. This movement pattern defined the scope of our analysis. That is, all safety, interaction, and performance evaluations in this study are focused on these left-turn movements, offering a consistent basis for comparing AV behavior in busy urban settings.

\section{Methodology}

\subsection{Interaction Detection}
We identified interactions by computing relative distances and angles between agents ($0^\circ$–$360^\circ$), assigning them to detection zones adapted from Tesla Vision’s field-of-view framework . These zones are used for illustrative analysis of behavioral differences and do not imply that the AVs in the dataset are Teslas. Key interaction metadata, including agent IDs and spatial relationships, were stored for further analysis.


\subsection{TTC and PET}
Time-to-Collision (TTC) was calculated using the relative motion between two interacting agents:
\begin{equation}
    TTC = \frac{d}{V_{\text{rel}}}
\end{equation}
where $d$ is the Euclidean distance between agents and $V_{\text{rel}}$ is the projected relative velocity along the line of motion. Special cases were handled to ensure stability:
\begin{itemize}
    \item If $V_{\text{rel}} \leq 0$: $\text{TTC} = \infty$ ( divergent or stationary). 
    \item If $d = 0$: $\text{TTC} = 0$ (overlap in space).
\end{itemize}

Post-Encroachment Time (PET) was defined as the absolute time difference between the moments when two agents reached their point of closest proximity, subject to a proximity threshold of 5 meters. PET was only computed for pairs whose minimum separation fell below this threshold, indicating potential spatial conflict.

\subsection{Time Headway Estimation}
We computed time headways for AVs and HDVs during entry and exit transitions based on lane mapping. Only headways under 5 seconds were retained to capture immediate following behavior, filtering out free-flow and signal-induced gaps.

\subsection{Platoon Identification via Affine Spacing}

We define platoon boundaries using the affine spacing policy, which accounts for variable vehicle dynamics and communication capabilities \cite{Michael_Shaham_2024}. A vehicle is considered part of the platoon if its spacing to the its leader satisfies:

\begin{equation}
d_{\text{actual}} - (d_0 + h v) \leq \epsilon
\end{equation}

where \( d_0 \) is the standstill distance, \( h \) is the desired time headway, and \( \epsilon \) is a tolerance term. We adopt \( h = 2.0 \, \text{s} \) and \( d_0 = 4 \, \text{m} \) based on observed behavior in urban HDV traffic \cite{Ahn2007,Nowell2000}, while allowing a tolerance of \( \epsilon = 5 \, \text{m} \) to accommodate noise in video-based trajectories. Vehicles violating this condition consecutively are excluded from the platoon.

This approach is applied to identify vehicles upstream of an AV entering the intersection during green at speeds exceeding 2 m/s, ensuring we capture only active, behaviorally-linked followers rather than queued vehicles at red.

\subsection{String Stability via L2 Norm Ratio}

String stability is evaluated by comparing acceleration magnitudes across consecutive vehicles in each platoon. For each vehicle \( n \), we extract its acceleration norm over a 10-second window centered at its entry time \( t_0 \):

\begin{equation}
G_n = \frac{\|a_n\|}{\|a_{n-1}\|}, \quad \text{where } a_n = \left( \sum_{t=t_0 - 5}^{t_0 + 5} a_n(t)^2 \right)^{1/2}
\end{equation}

String stability holds if $G_n \leq 1$.
The 10-second window ensures that interactions are captured during the period in which vehicles are behaviorally impacted by one another; beyond either bound of the [-5,5] interval, divergent trajectories and signal phases render comparisons behaviorally invalid.

\subsection{Pedestrian Hesitation in Presence of Turning Vehicles}

To identify instances of hesitation during vehicle-pedestrian interactions, we applied a pattern-based search across pedestrian trajectories. Hesitation was defined as a temporal pattern where a pedestrian initially moved forward (\(v > 0.5 \, \text{m/s}\)), slowed down for several frames, then picked up speed again, all while a left-turning vehicle was present within 15 meters. 

\section{Results}

\subsection{TTC Patterns and Conflict Exposure}

\begin{figure}[H]
    \centering
    \includegraphics[width=0.98\linewidth]{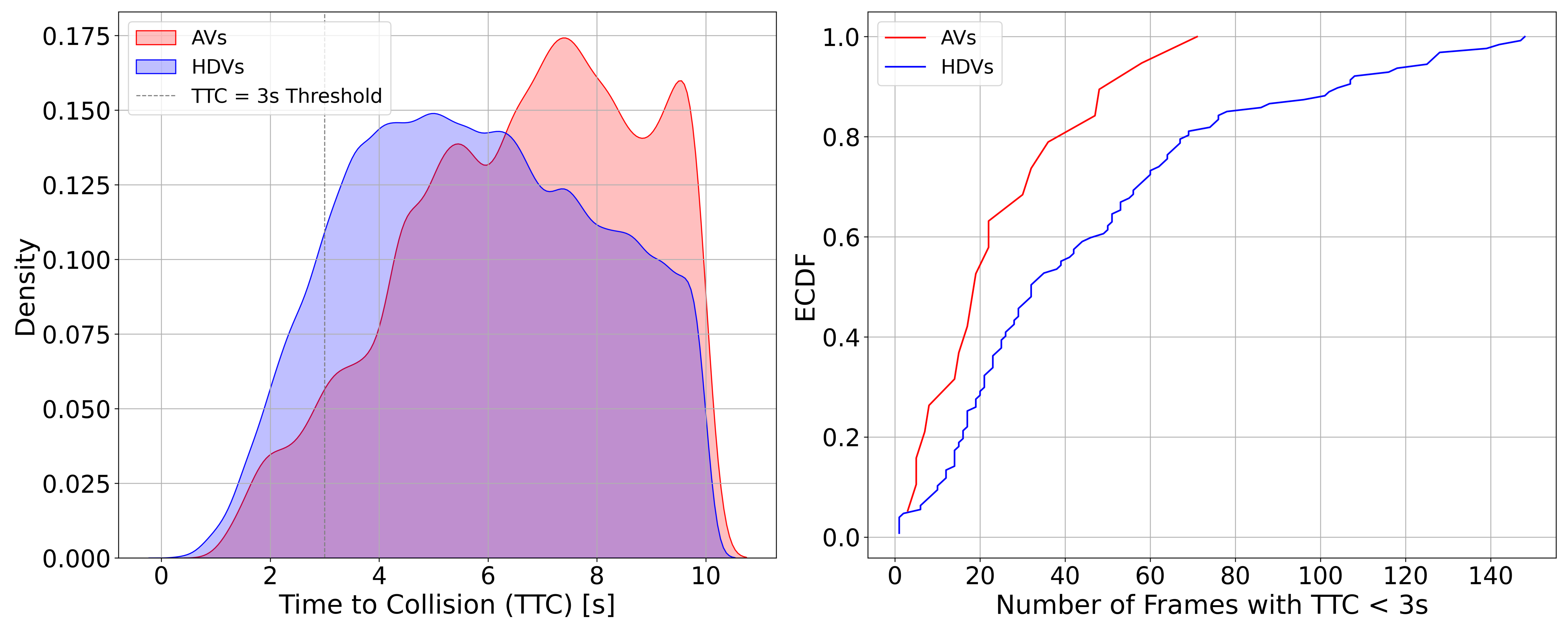}
    \caption{Left: Kernel Density Estimates (KDE) of TTC values for AVs and HDVs interacting with VRUs, with a 3s threshold shown. Right: Empirical Cumulative Distribution Function (ECDF) of total time spent with TTC $<$ 3s.}
    \label{fig:ttc_combined}
\end{figure}

Figure~\ref{fig:ttc_combined} provides a combined view of Time-to-Collision (TTC) behavior. 

The left panel shows that AVs consistently maintain higher TTC values, reinforcing their more conservative, anticipatory approach to VRU interactions. HDVs, by contrast, more frequently operate near the 3-second threshold. The right panel reveals that AVs more commonly experience brief episodes of TTC $<$ 3s, reflected by a steeper early ECDF rise. HDVs are less frequently exposed to these short TTCs, but when they are, the exposure tends to last longer, implying slower conflict resolution and prolonged proximity risk.

\subsection{TTC by Detection Zone Results}

\begin{figure}[H]
    \centering
    \includegraphics[width=0.9\linewidth]{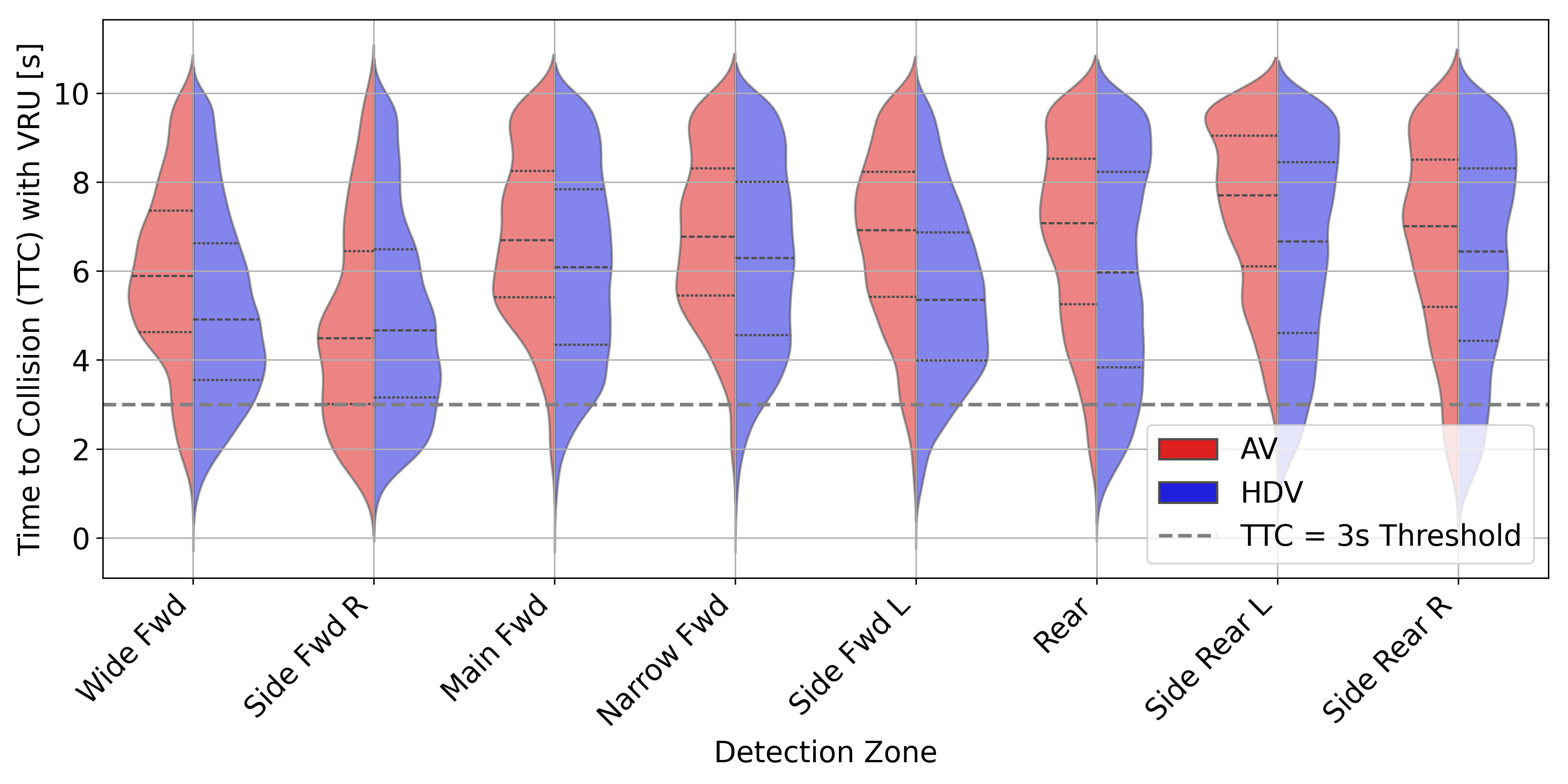}
    \caption{Distribution of TTC with VRUs by Detection Zone.}
    \label{fig:ttc_violin_zone}
\end{figure}

Figure~\ref{fig:ttc_violin_zone} presents the TTC distributions for AVs and HDVs interacting with VRUs across Tesla Vision detection zones, based on filtered TTC values below 10 seconds and zone-specific interaction mapping.

AVs consistently demonstrate higher TTC values in forward-facing zones (\textit{Main Fwd, Narrow Fwd, Wide Fwd}), indicating more anticipatory or smoother responses when a VRU is detected in the direction of motion. This behavior is especially evident in the \textit{Side Fwd L} zone, where AVs show a clear safety advantage (TTCs well above 3s), whereas HDVs show greater risk in short-range interactions.

Interestingly, HDVs slightly outperform AVs in the \textit{Side Fwd R} zone, likely due to conservative behavior on the off-turning side, though this zone plays a marginal role in left-turn pedestrian interactions. Rear zones are visually retained but not interpreted, as no meaningful vehicle-VRU interaction occurs in the rear zones at these locations.

Overall, these results confirm that AVs respond more conservatively in zones directly aligned with the left-turn path, while HDVs display faster clearance but occasionally shorter TTCs, especially at peripheral angles.

\subsection{PET Patterns and Clearance Behavior}

\begin{figure}[H]
    \centering
    \includegraphics[width=0.98\linewidth]{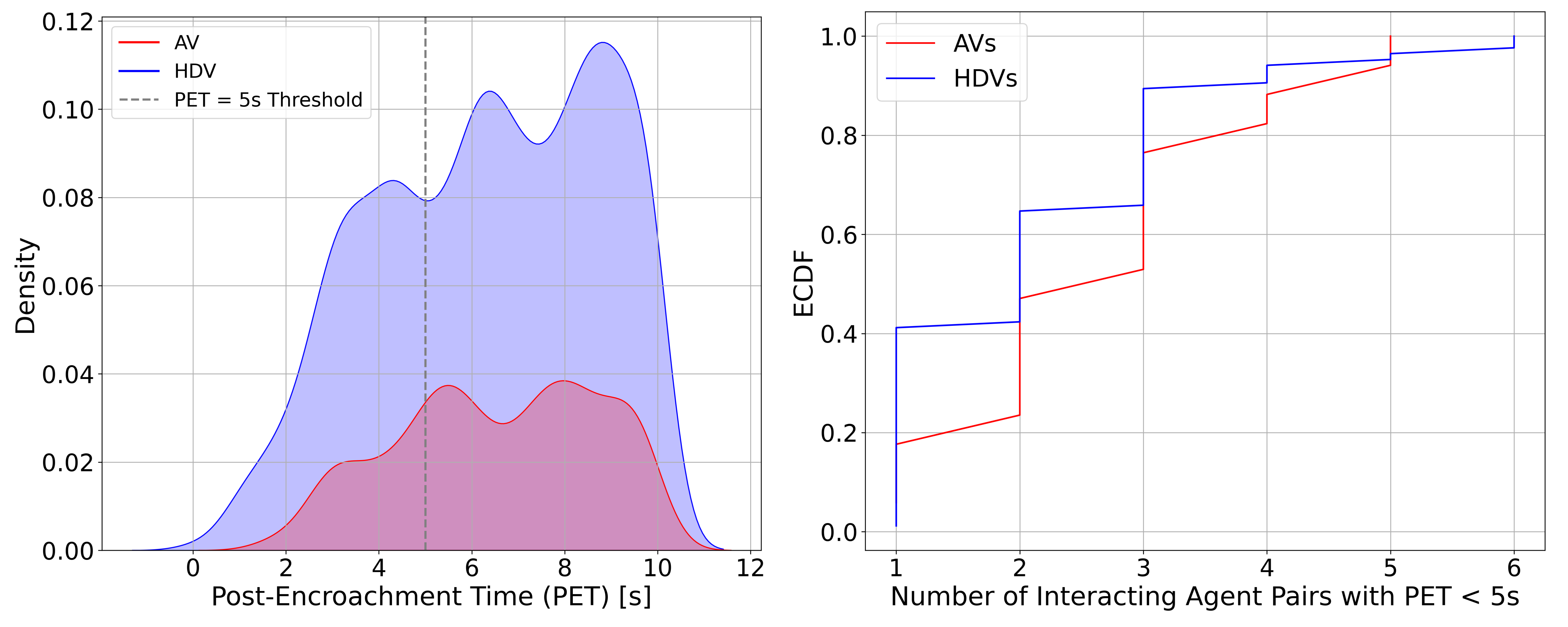}
    \caption{Left: KDE of PET values for AVs and HDVs interacting with VRUs. Right: ECDF of the number of PET $<$ 5s interactions per vehicle.}
    \label{fig:pet_combined}
\end{figure}

Figure~\ref{fig:pet_combined} presents a combined view of PET behavior.  While AVs maintain higher minimum TTC values than HDVs, the PET distributions reveal the opposite trend. 

AVs experience a larger number of PET $<$ 5s interactions per vehicle compared to HDVs, as shown by the steeper early rise of the ECDF curve. In contrast, HDVs generally achieve longer clearance times, reflected by higher PET values in the left panel. These contrasting patterns suggest that although AVs react earlier to potential conflicts, they often operate with tighter spatial or temporal proximity during conflict clearance compared to human drivers.

\subsection{Deceleration Patterns During Left Turns}

We analyzed deceleration behavior for AVs and HDVs at left-turn locations where interactions with VRUs are expected during permitted left turn.

\begin{figure}[H]
    \centering
    \includegraphics[width=0.98\linewidth]{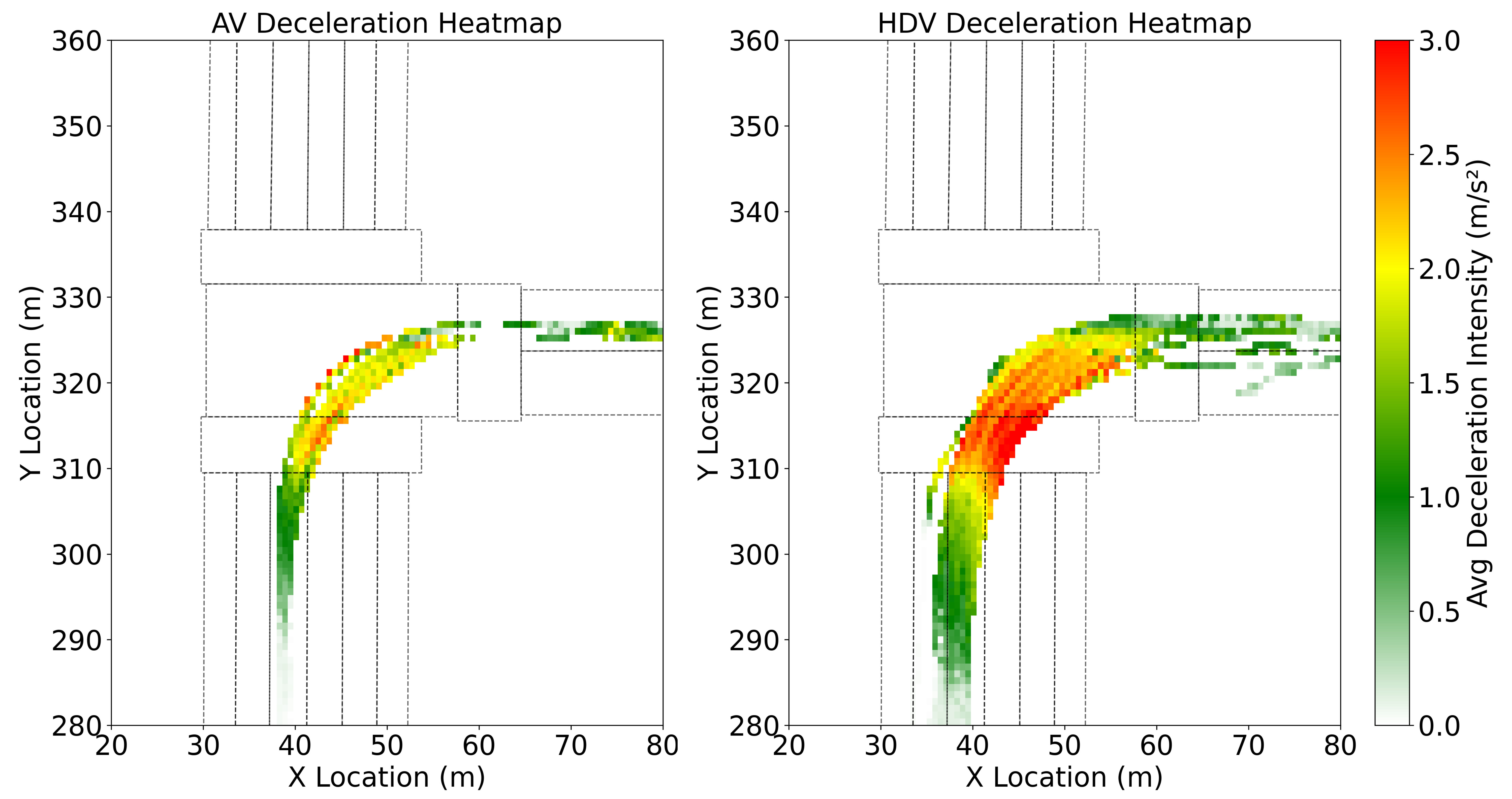}
    \caption{Deceleration heatmap for AVs (left) and HDVs (right) during a selected left-turn approach.}
    \label{fig:heatmap_av_hdv}
\end{figure}

Figure~\ref{fig:heatmap_av_hdv} shows the difference in the deceleration intensity heatmaps (scaled at 0–3 m/s²) between AVs and HDvs. As shown in the figure, AVs concentrated braking along the arc of the turn with smoother, anticipatory deceleration, while HDVs exhibited more intense and spatially scattered braking—often near pedestrian conflict zones. These patterns suggest a contrast between proactive and reactive behavioral strategies.

\begin{figure}[H]
    \centering
    \includegraphics[width=0.9\linewidth]{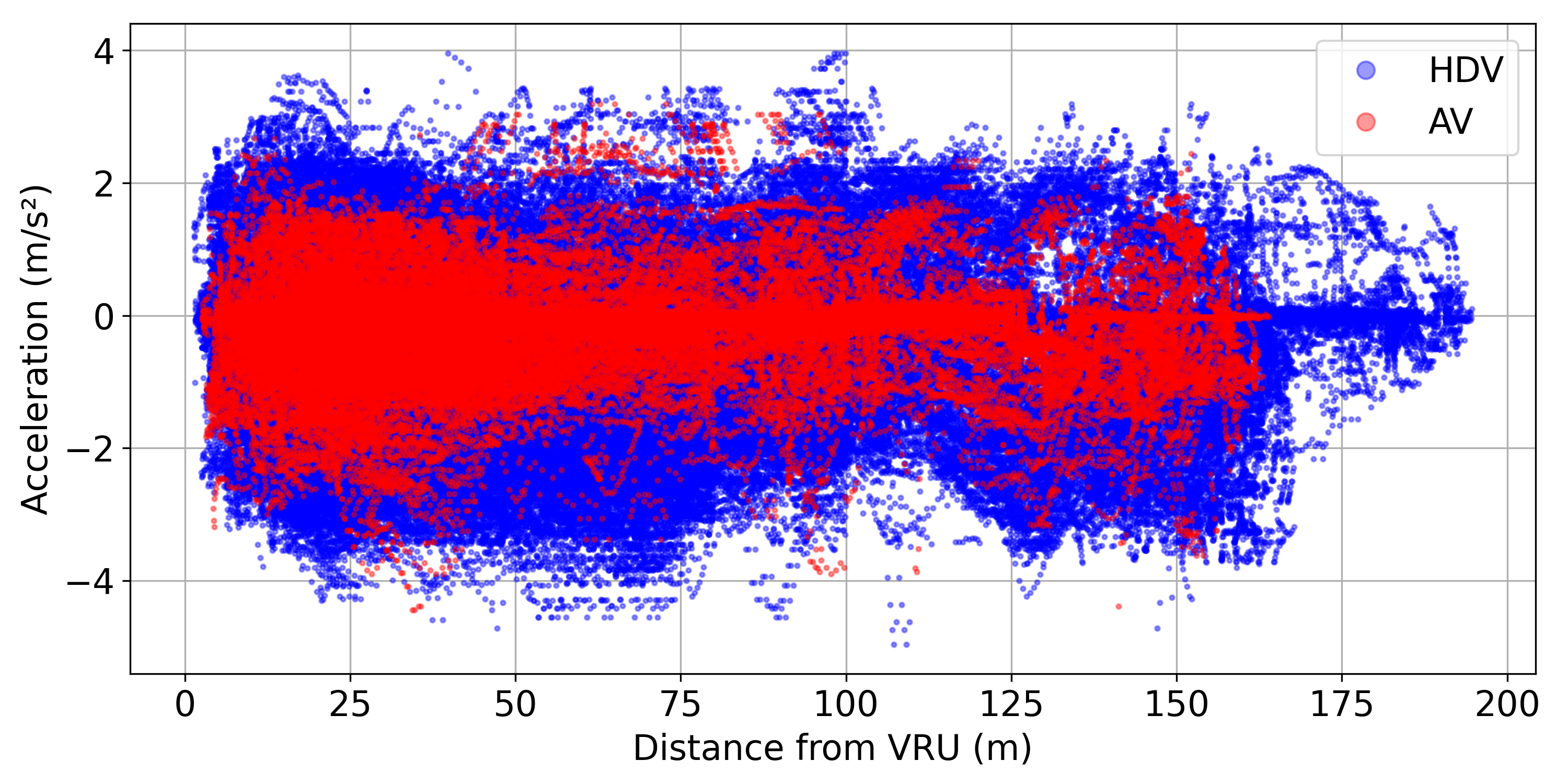}
    \caption{Acceleration vs Distance to VRUs for AV and HDV interactions during left turns.}
    \label{fig:accel_distance_scatter}
\end{figure}

To further quantify these trends, we examined acceleration values relative to the distance between vehicles and VRUs. Figure~\ref{fig:accel_distance_scatter} shows that AVs generally began slowing earlier than HDVs, indicating a more defensive posture in managing crosswalk interactions during turning.

\subsection{Headway Patterns for Entry and Exit Transitions}
\begin{figure}[H]
    \centering
    \includegraphics[width=0.9\linewidth]{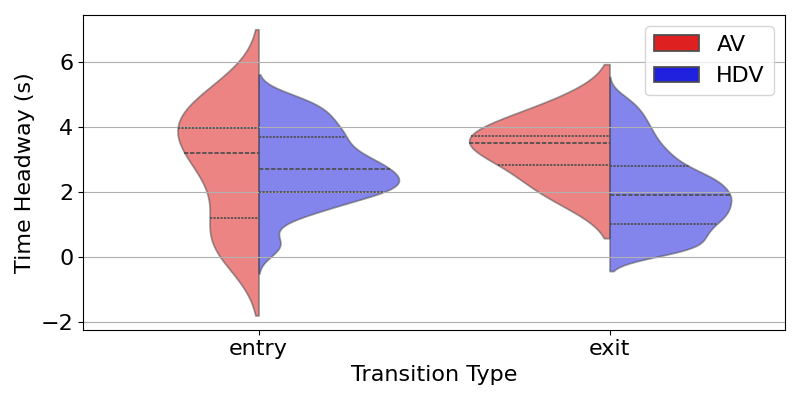}
    \caption{Violin plot of AV and HDV headways during entry and exit transitions.}
    \label{fig:violin_headways}
\end{figure}

Figure~\ref{fig:violin_headways} compare the distributions of AV and HDV time headways during entry and exit transitions. Entry headways appear similar across vehicle types, reflecting signal or queue constraints. Exit headways show greater separation, with AVs maintaining higher buffers.

\subsection{String Stability for First Followers}

Due to the predominance of 2-vehicle platoons in the dataset and the limited number of AVs available, our comparison of string stability between AV and HDV led platoons focuses specifically on the first follower. This simplification allows for a consistent comparison without sample imbalance or propagation artifacts from longer platoon chains.

\begin{figure}[H]
    \centering
    \includegraphics[width=0.9\linewidth]{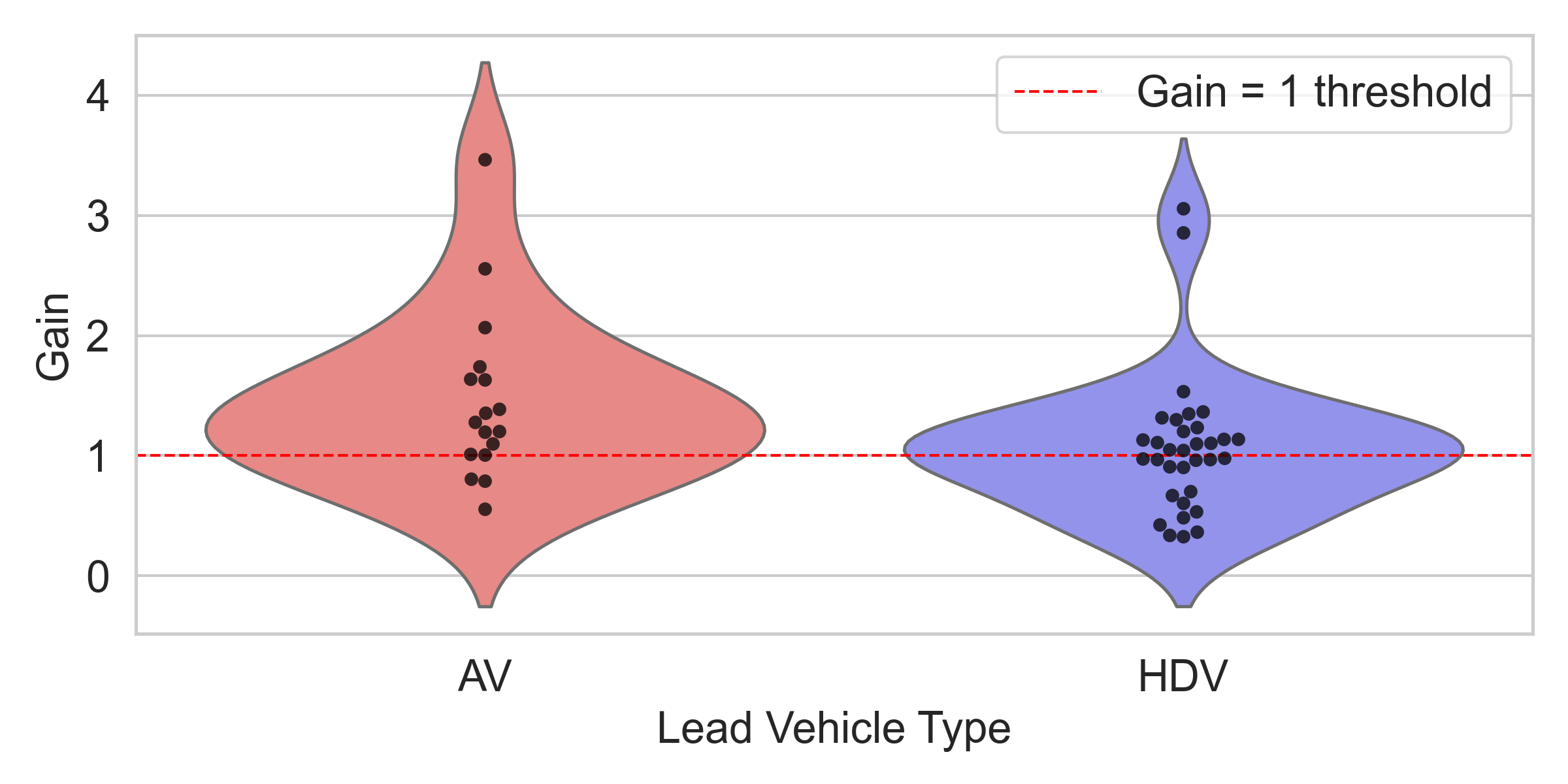}
    \caption{Distribution of gain values for first followers in AV and HDV led platoons. Gains above 1 indicate amplification of acceleration disturbances.}
    \label{fig:first_follower_violin}
\end{figure}

Figure~\ref{fig:first_follower_violin} illustrates the distribution of gain values (defined as the L2 norm ratio between the follower's and leader's acceleration signals) for position-1 vehicles under AV and HDV led platoons.  

Notably, HDV-led platoons exhibit slightly lower and more concentrated gain distributions, indicating better attenuation of acceleration disturbances by the first follower.

One possible explanation stems from the operational context of the TGSIM Foggy Bottom dataset: a dense, signalized urban environment with frequent pedestrian activity, closely spaced intersections, and mixed traffic. AVs navigating this environment may adopt overly conservative or reactive driving behaviors. This manifests itself in abrupt deceleration, extra spacing near VRUs, or hesitation at occluded intersections, which can introduce small fluctuations in acceleration profiles. These small disturbances, even if minor, may be misinterpreted or exaggerated by the first follower, particularly in cases where AV behavior diverges from traditional human-following norms. In contrast, HDV-led platoons may benefit from more consistent driver expectations—that is, human followers can more easily anticipate the behavior of human leaders, especially in familiar environments like traffic lights and crosswalks.

\subsection{Pedestrian Hesitation During Left Turns}

\begin{figure}[H]
\centering
\includegraphics[width=0.99\linewidth]{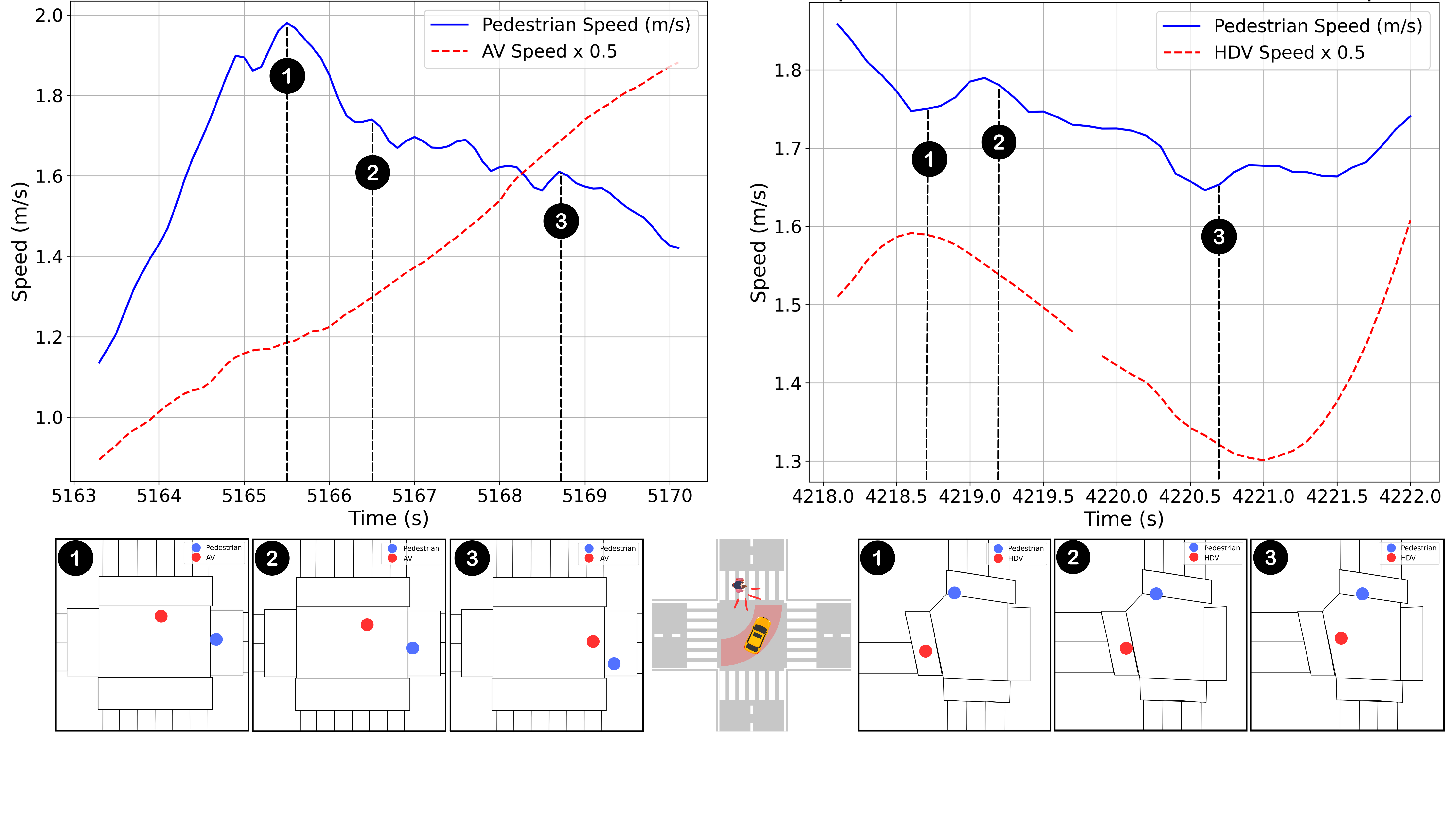}
\caption{Examples of pedestrian hesitation in interactions with AVs (left) and HDVs (right). Numbered markers indicate key interaction moments, with corresponding layouts shown below.}
\label{fig:hesitation_results}
\end{figure}

We extracted 23 hesitation cases in total: 14 involving HDVs and 9 involving AVs. Figure~\ref{fig:hesitation_results} compares representative speed profiles of pedestrians in HDV (right) and AV (left) encounters. Each panel shows the pedestrian’s speed over time, overlaid with the scaled speed of the nearby turning vehicle, and includes snapshots at key interaction points.

Interestingly, while AVs tend to maintain a stable and smoother speed profile when navigating intersections, as shown earlier in this study, this consistency may not always be perceived as reassuring by pedestrians. In several hesitation cases, pedestrians began slowing down just as the AV entered the intersection and continued to do so even without sudden AV movements. The lack of visible deceleration from the AV may have signaled a lack of yielding intent. In contrast, HDVs often slowed down earlier as they approached a crossing pedestrian. This behavior, particularly when occurring near the stop line, may have served as a more visible cue of yielding, prompting pedestrians to resume movement more confidently.

Although fewer hesitation cases were observed with AVs overall, the fact that 9 cases were recorded, despite the relatively small number of AVs in the dataset, raises concerns about how pedestrians respond to AVs during left turn maneuvers.

\subsection{AV–VRU Co-Occupancy in Pedestrian Zones}
We analyzed whether automated vehicles (AVs) entered pedestrian zones while VRUs (pedestrians, cyclists, or scooters) were actively crossing (\(v > 0.5 \, \text{m/s}\)). Across four crosswalk zones, 15 co-occupancy cases involving AVs were identified. While human drivers often initiate permissive left turns once VRUs clear their path, AVs are held to stricter safety expectations. Sharing crosswalk zones with crossing VRUs raises concerns about AV compliance with right-of-way laws, which generally require drivers to yield until VRUs fully exit. These findings highlight challenges in adapting AV behavior to varied legal and social norms across jurisdictions.

\section{Discussion}

The multifaceted evaluation across safety, interaction, and traffic performance dimensions reveals nuanced distinctions in how AVs and HDVs navigate urban left-turns maneuvers, summarized in Table~\ref{tab:dim_metric_summary}.

\begin{table}[H]
\caption{Summary of Metrics Across Dimensions}
\label{tab:dim_metric_summary}
\scriptsize
\renewcommand{\arraystretch}{1.3}
\begin{tabularx}{\linewidth}{p{1.1cm} p{1.5cm} X}
\toprule
\textbf{Dimension} & \textbf{Metric} & \textbf{Interpretation of Findings} \\
\midrule

Safety & TTC & AVs maintained longer TTCs near VRUs, indicating more anticipatory behavior. \\[0.5ex]

Safety & PET & AVs showed more PET under 5s per vehicle, suggesting close but cautious interactions. \\[0.5ex]

Safety & Co-occupancy & AVs co-occupied crosswalks with VRUs less often, but even rare events are concerning due to higher safety expectations. \\[0.5ex]

Interaction & Hesitation & Pedestrians slowed near AVs despite smooth motion, likely due to unclear yielding cues. \\[0.5ex]

Interaction & Deceleration & AVs decelerated smoothly and predictably; HDVs braked more abruptly and variably. \\[0.5ex]

Performance & Headway & AVs maintained larger exit headways, reflecting caution. HDVs followed more tightly, improving flow. \\[0.5ex]

Performance & String Stability & HDV-led platoons showed lower acceleration gain values, indicating better disturbance attenuation. \\

\bottomrule
\end{tabularx}
\end{table}

AVs consistently exhibited behavior aligned with safety prioritization. They maintained higher TTC values and exhibited smoother deceleration patterns near pedestrian zones. These results are consistent with prior work showing that AVs adopt conservative braking and maintain larger safety margins in complex environments \cite{Saeed_Rahmani_2024, Sunny_Singh_2023, Zou2023-dn}. 

However, when looking at PET on a per-vehicle basis, AVs were involved in more interactions with PET values below 5 seconds. This suggests that AVs may be clearing crosswalks more closely behind VRUs compared to HDVs. Although these close clearances did not result in conflicts, their frequency raises concerns about how AVs judge safe separation. It may reflect a reliance on internal models that assume pedestrians will behave predictably, which does not always hold true in busy or ambiguous settings \cite{Rezwana2025-pt, Lanzaro2023-gl}.

This misalignment becomes more visible in the pedestrian hesitation analysis. Despite AVs maintaining gradual speeds, pedestrians were observed slowing down when encountering AVs, likely due to a lack of visible yielding cues. HDVs, in contrast, demonstrated more pronounced deceleration at key moments, which may have offered more reassuring signals to crossing pedestrians. The literature similarly points to AVs’ lack of adaptive social signaling, such as informal gestures or dynamic feedback, as a source of pedestrian uncertainty during interactions \cite{Izquierdo2024-yu, Lau2024-ex}.

From a traffic performance perspective, HDVs demonstrated advantages in terms of shorter headways and more stable acceleration propagation in platoons. AVs tend to maintain larger spacing, particularly during intersection exits, reducing throughput and contributing to disturbance amplification in follower vehicles. While this behavior can be interpreted as more safety-conscious, it did not translate into better string stability outcomes. In fact, HDV-led platoons demonstrated reduced propagation of acceleration perturbations, as supported by our string stability analysis. This finding aligns with work highlighting the occasional advantage of human adaptability in maintaining flow stability under pressure \cite{Das2024-tr}.

A more critical safety concern emerges in the co-occupancy analysis. While HDVs were more frequently present in pedestrian zones while VRUs were crossing, AVs were also found to co-occupy crosswalk zones in 15 distinct cases. Given the stricter safety expectations for AVs, this raises flags about either perception limitations or edge-case decision policies that warrant reevaluation.

These observations reinforce our conceptual framework (Figure~\ref{fig:flowchart_framework}), which highlights potential trade-offs between dimensions. AVs may excel in surrogate safety metrics but fall short on interaction clarity or traffic efficiency. Conversely, HDVs may communicate intent more effectively and maintain better flow, albeit with more variable and potentially riskier behavior.

In this work, we posit that AV operation should aim for consensus-aware strategies, those that simultaneously satisfy all 3 dimensions. We adopted a \textit{consensus zone mapping} approach to identify time-space segments where AV behavior concurrently satisfied safety, interaction, and performance objectives. For safety, we defined consensus-aligned instances as those with $TTC > 3\,\text{s}$ and $PET > 5\,\text{s}$; for interaction quality, we looked for episodes with no detected pedestrian hesitation and stable pedestrian speed; for traffic performance, we required exit headways below 4 seconds and a string stability gain $G \leq 1$ for the first follower. Out of all AV–VRU interaction frames, only 1.63\% met all three consensus conditions simultaneously, 27.65\% satisfied two conditions, and 70.71\% met at most one or none, indicating that full consensus across safety, interaction, and performance remains rare.Importantly, not all dimensions are equally tractable for AVs. Safety alignment was the most commonly observed. Interaction quality appeared partially achievable. Traffic performance, however, remained the least aligned with AV behavior. We show a directional interaction matrix (Table~\ref{tab:metric_interactions}) highlighting how changes in one metric affect safety, interaction quality, and traffic performance.


\begin{table}[H]
\caption{Metric Interactions Across Dimensions}
\label{tab:metric_interactions}
\scriptsize
\renewcommand{\arraystretch}{1.3}
\setlength{\tabcolsep}{2pt}
\centering
\begin{tabular}{p{1.9cm} p{1.4cm} p{1.4cm} p{1.7cm} p{1.7cm}}
\toprule
\textbf{Metric} & \textbf{Primary\newline Dimension} & \textbf{Effect on\newline Safety} & \textbf{Effect on\newline Interaction} & \textbf{Effect on\newline Performance} \\
\midrule
TTC ↑             & Safety        & --                     & Hesitation ↓            & Headway ↑            \\
PET ↑             & Safety        & --                     & Hesitation ↓            & Throughput ↓         \\
Co-occupancy ↑    & Safety        & --                     & Hesitation ↓            & Clearance ↓          \\
Hesitation ↑      & Interaction   & TTC ↓                 & --                      & Stability ↓          \\
Deceleration ↑    & Interaction   & PET ↑                 & --                      & Stability ↓          \\
Headway ↑         & Performance   & TTC ↓                 & Hesitation ↑           & --                   \\
String Stability ↑ & Performance  & PET ↑                & Deceleration ↑          & --                   \\
\bottomrule
\end{tabular}
\end{table}

\section{Conclusion and Future Work}

This study provided a multi-dimensional evaluation of AV behavior in mixed traffic, focusing on surrogate safety measures, pedestrian interaction quality, and traffic performance indicators. Using high-resolution trajectory data, we analyzed key metrics such as TTC, PET, co-occupancy, pedestrian hesitation, deceleration behavior, headways, and string stability to characterize differences between AVs and HDVs. 

Our findings revealed that while AVs often exhibit behaviors aligned with safety, such as smooth deceleration and larger spacing, these same behaviors can reduce interaction clarity and degrade traffic performance. In contrast, HDVs demonstrated more variable but sometimes more intuitive and traffic flow positive behavior. To reconcile these tensions, we introduced a consensus framework to identify how safety, interaction, and performance goals interact and co-occur. 

We hope our endeavors will spur interest and motivate further work on designing consensus-aware AV models, for optimal real-world deployment.


\bibliographystyle{IEEEtran}
\bibliography{references}

\end{document}